\documentclass[prb,twocolumn,showemail,groupedaddress,showpacs]{revtex4-2}
\usepackage{physics}
\usepackage[export]{adjustbox}
\usepackage{amsmath}
\usepackage{amssymb}
\usepackage[normalem]{ulem}

\usepackage[unicode=true,pdfusetitle,
 bookmarks=true,bookmarksnumbered=false,bookmarksopen=false,
 breaklinks=false,pdfborder={0 0 0},backref=false,colorlinks=true,citecolor=blue,urlcolor=violet 
]{hyperref}
\usepackage{xcolor}
\usepackage{graphicx}
\usepackage{orcidlink}
\newcommand{\obs}{\boldsymbol}
\newcommand{\vect}{\vec}
\renewcommand{\i}{\mathrm{i}}
\newcommand{\unity}{1\kern-0.25em\text{l}}

\begin{document}
\title{Unifying Quantum and Classical Dynamics}

\author{Abdul Rahaman Shaikh\,\orcidlink{0000-0003-2295-027X}}
\email{abdulrahaman@ctp-jamia.res.in}
\affiliation{Centre for Theoretical Physics, Jamia Millia Islamia, New Delhi-110025, India}
\author{Tabish Qureshi\,\orcidlink{0000-0002-8452-1078}}
\email{tabish@ctp-jamia.res.in}
\affiliation{Centre for Theoretical Physics, Jamia Millia Islamia, New Delhi-110025, India}

\begin{abstract}
Classical and quantum physics represent two distinct theories; however, quantum physics is regarded as the more fundamental of the two. It is posited that classical {mechanics} should arise from quantum mechanics under certain limiting conditions. Nevertheless, this remains a challenging objective. In this work, we explore the potential for unifying the dynamics of classical and quantum physics. This discussion does not suggest that classical behavior emerges from quantum mechanics; rather, it demonstrates the exact equivalence between the dynamics of quantum observables and their classical counterparts. It is shown that the Heisenberg equations of motion can be cast in a form that is structurally equivalent to Newton's equations of motion, with $\hbar$ dropping out from the equations. In a generalized analysis, the Heisenberg equations are cast in a form that is identical to the classical Hamilton's equations of motion.
{This implies that both quantum and classical dynamics are governed by the same form of equations, with the Heisenberg operators substituting the classical observables}.
\end{abstract}

\maketitle
\section{Introduction}
Quantum mechanics represents an exceptionally successful framework for elucidating the behavior of microscopic particles. This theory has not only been effectively utilized in a wide array of physical phenomena, but it has also undergone rigorous testing through precise experimental methods. Despite being primarily developed through intuition rather than strict derivation, it has demonstrated remarkable accuracy and has not required any modifications. Given its effective description of microscopic objects, there is a prevailing belief that it should similarly be capable of describing macroscopic objects, which are after all composed of these microscopic particles. The dynamics of macroscopic objects in non-relativistic domain, is described well by classical mechanics, which may be formulated equivalently in terms of Newton’s laws, or in the Hamiltonian formalism. This naturally leads one to the expectation that classical mechanics should emerge out of quantum mechanics. The mechanism by which it may happen is under active debate with widely divergent views. At the most modest end is the \emph{correspondence principle} proposed by Niels Bohr \cite{CP1}, which may be interpreted as classical behavior of quantum systems emerging in the limit of large quantum numbers. Another approach is the Wentzel–Kramers–Brillouin approximation, or the WKB method in which one imagines a semi-classical scenario where $\hbar$ is considered a small parameter \cite{WKB}. In this approach, classical regime would correspond to assuming $\hbar$ to be negligibly small {\cite{Klein2012}. In fact, Dirac states that
“\dots classical mechanics may be regarded as the limiting case of quantum mechanics when $\hbar$ tends to zero \cite{Dirac}.”}
Similarly, in the Feynman path-integral formulation of quantum mechanics~\cite{FeynmanHibbs}, the classical principle of least action emerges in the $\hbar \to 0$ limit due to the rapid oscillation of the phase $e^{iS/\hbar}$, which suppresses non-stationary paths and selects trajectories satisfying $\delta S = 0$, corresponding to the classical path. There are other approaches, based on Ehrenfest theorem, for obtaining classical dynamics in the limit $\hbar\to 0$ \cite{Kazandjian}. At the radical end are ideas that quantum theory needs \emph{nonlinear} modification in order to fully recover classical mechanics \cite{collapse,gravity}.  Somewhere in between are ideas like decoherence where classical behavior is an outcome of weak interaction of quantum systems with certain environment degrees of freedom \cite{dec}. There is also a diametrically opposite view according to which quantum mechanics is not a fundamental theory, but rather an emergent phenomenon \cite{adler}.

Before we are anywhere near answering the above questions, we may try to explore if there is some common ground between the two theories. Identifying such a common ground may be helpful in finally finding the solution to the problem. More specifically we address the issue of dynamics of observable quantities in the two theories. In classical mechanics, the dynamics of observables is governed by classical equations of motion, such as Newton’s equations or their Hamiltonian formulation. In quantum mechanics, there exist several equivalent formulations to describe dynamics, the most popular being the Schr\"odinger and Heisenberg formalisms. The  Schr\"odinger equation governs the time evolution of the quantum state, more commonly called the wavefunction. On the other hand the Heisenberg equation of motion governs the time evolution of operators which represent the observable quantities. Since the wavefunction doesn't have a classical counterpart, it seems natural to look for a connection between the Heisenberg equation of motion and the Newton's equation or Hamiltonian equation of motion. 

\section{Heisenberg picture}
To start with a simple case, we consider a particle of mass $m$, moving in {three dimensions}, in a potential $V(\vect{x})$. {Throughout this paper, boldface quantities represent operators, ordinary symbols represent scalars, arrows indicate vectors, and boldface symbols with arrows represent vector operators.} The Hamiltonian, which is the operator for energy, is then given by
\begin{equation}
\obs{H} = \frac{\vect{\obs{p}}^2}{2m} + V(\vect{\obs{x}}) ,
\end{equation}
{where $\vect{\obs{x}},\vect{\obs{p}}$ are the position and momentum vector operators}, respectively. The Heisenberg equation of motion for the $i^{th}$ component of position can be written as
\begin{equation}
\frac{d}{dt}\obs{x}_i(t) = \frac{\i}{\hbar}[\obs{H},\obs{x}_i(t)] ,
\end{equation}
where $\obs{x}_i(t) = \obs{U}_t\obs{x}_i(0)\obs{U}^\dag_t$,
$\obs{U}_t=\exp(\i \obs{H}t/\hbar)$ being the unitary time evolution operator {and $i$ is 1, 2 and 3 for $x$, $y$ and $z$-components, respectively}. Since $\obs{U}_t$ commutes with $\obs{H}$, and $\obs{x}_i$ commutes with $V(\vect{\obs{x}})$,
the above equation yields
\begin{equation}
m \frac{d}{dt}\obs{x}_i(t) = \obs{p}_i(t).
\label{Nx}
\end{equation}
Similarly, the Heisenberg equation of motion for the {$i^{th}$ component of the} momentum operator can be written as
\begin{equation}
\frac{d}{dt}\obs{p}_i(t) = \frac{\i}{\hbar}[\obs{H},\obs{p}_i(t)]
= \frac{\i}{\hbar}\obs{U}_t[V(\vect{\obs{x}}),\obs{p}_i(0)]\obs{U}^\dag_t .
\label{Hp}
\end{equation}
Now let us assume that the potential $V$ can be expanded in a Taylor series in any one component, say, $x_i$ around $x_{i0}$
\begin{eqnarray}
V(\vect{\obs{x}}) &=& \sum_{n=0}^{\infty} \frac{1}{n!}
\left.\frac{\partial^n V(\vect{\obs{x}})}{\partial \obs{x}_i^n}\right|_{\obs{x}_i=x_{i0}\unity} \left(\obs{x}_i-x_{i0}\unity\right)^n \nonumber\\
 &=& \sum_{n=0}^{\infty}
\frac{V_{x_{i0}}^{(n)}}{n!} \left(\obs{x}_i-x_{i0}\unity\right)^n ,
\end{eqnarray}
{where $V_{x_{i0}}^{(n)} \equiv
\left. \frac{\partial^n V(\vect{\obs{x}})}{\partial \obs{x}_i^n} \right|_{\obs{x}_i=x_{i0}\unity}$ and $\unity$ denotes the identity operator.
Here, $V_{x_{i0}}^{(n)}$ denotes the $n$th partial derivative of
$V(\vect{\obs{x}})$ with respect to $\obs{x}_i$, evaluated at
$\obs{x}_i=x_{i0}\unity$, while retaining its dependence on the
remaining components $\obs{x}_{j\neq i}$}.
{A derivative such as $\frac{\partial \obs{V}}{\partial \obs{x}_i}$ should be interpreted as first treating $\obs{V}$ as a classical function of the variable $x_i$, differentiating it with respect to $x_i$, and then promoting $x_i$ to the quantum operator $\obs{x}_i$ in the resulting expression. While this serves as a practical computational recipe, it is supported by a rigorous theory of derivatives of operator-valued functions \cite{Suzuki1999}. Another point to note here is that $V_{x_{i0}}^{(n)}$ is effectively treated as a scalar function, as far as the Hilbert space of $\obs{x_i}$ is concerned, because whatever be its functional form, the end result will be a scalar function multiplied by the identity operator. The identity operator anyway gets absorbed in the term $(\obs{x}_i-x_{i0}\unity)^n$.}
Using the series form of the potential, the commutator of $\obs{p}_i$ with the potential can be written as \cite{Sakurai}
\begin{eqnarray}
[V(\vect{\obs{x}}),\obs{p}_i(0)] &=& \sum_{n=0}^\infty
\tfrac{1}{n!} V^{(n)}_{x_{i0}}[(\obs{x}_i-x_{i0}\unity)^n,\obs{p}_i]\nonumber\\
&=& \i \hbar\sum_{n=0}^\infty
\tfrac{1}{n!} V^{(n)}_{x_{i0}}n(\obs{x}_i-x_{i0}\unity)^{n-1} \nonumber\\
&=& \i\hbar V'_{x_i}(\vect{\obs{x}}) ,
\end{eqnarray}
{where $V'_{x_i}(\vect{\obs{x}})$ represents the derivative of the potential with respect to $\obs{x}_i$, and} we have used the well known commutation relation
$[\obs{x}_i^n,\obs{p}_i]= \i \hbar n \obs{x}_i^{n-1}$. Then the Heisenberg equation of motion for $\obs{p}_i$ (\ref{Hp}) assumes the form
\begin{equation}
\frac{d}{dt}\obs{p}_i(t) = - V'_{x_i}(\vect{\obs{x}}(t)) .
\label{Np}
\end{equation}
Using (\ref{Nx}) and (\ref{Np}), one can write
\begin{equation}
m\frac{d^2}{dt^2}\obs{x}_i(t) = \frac{d}{dt}\obs{p}_i(t) = - V'_{x_i}(\vect{\obs{x}}(t)) .
\label{Nxp}
\end{equation}
This is a fully quantum mechanical equation involving operators, but has the
exact form of the Newton's equation of motion. The only assumption that we made
in arriving at this equation is that the potential can be expanded in a Taylor series. Remarkable is the fact that although (\ref{Nxp}) is a fully quantum mechanical equation, it does not contain $\hbar$. Unlike the WKB method, one doesn't need to make any approximation like ``$\hbar$ small'' in order to establish the quantum-classical connection. {However, this should not be interpreted as implying that quantum nature somehow disappears from the formalism. Quantum nature is very much there, tacit in the ubiquitous commutation relations. It is just that $\hbar$ is not what distinguishes quantum evolution equations from corresponding classical equations, as is widely believed \cite{Klein2012,Kazandjian}}.

It may be instructive to look at a couple of specific examples. For a {one dimensional} harmonic oscillator with potential energy $V(\obs x)=\frac{1}{2}m\omega^2 \obs x^2$, the Heisenberg equation looks like
\begin{equation}
m\frac{d^2}{dt^2}\obs{x}(t) = -m\omega^2\obs{x}(t) ,
\end{equation}
which is the same as Newton's equation for a {harmonic} oscillator. It is easy to see that generalization to many particles is straightforward, simply because observables of different particles commute with each other. As an example, we focus on a small mass $m$ that is orbiting around a larger mass $M$. We assume that $M$ is significantly greater than $m$, which allows us to position the center of mass of the system at the center of the larger mass, designated as the origin. The Hamiltonian is given by
$$ \obs{H} = \frac{\vect{\obs{P}}^2}{2M} + \frac{\vect{\obs{p}}^2}{2m} - \frac{G mM}{|\vect{\obs{r}}|} ,$$
where $\vect{\obs{r}}$ is the three-dimensional position operator of the lighter particle in the center-of-mass frame  {and $\vect{\obs{P}}$, $\vect{\obs{p}}$ are the momentum of larger mass and smaller mass in the center of mass frame, respectively}. Following the procedure described above, the Heisenberg equation for the momentum operator of the smaller particle is given by
{\begin{equation}
\frac{d}{dt}\vect{\obs{p}}(t) = -\frac{G mM}{|\vect{\obs{r}}|^2} \hat{r},
\end{equation} 
where $\hat{r}$ is a unit vector in the direction of $\vect{\obs{r}}$.}
Again we observe that the Heisenberg equation coincides with the Newton's equation.

{One would notice that the preceding derivations rely on {the Taylor expandability} of the potentials. This would naturally exclude situations where we have discontinuous potentials,
delta-function barriers, or hard-wall boundaries. However, we may argue that these cases are mathematical representations of real potentials which are not strictly discontinuous. One may say, for quantum particles, there are no strictly hard walls. Irrespective of what one may conclude, this limitation of the procedure should be kept in mind.}


\section{Charged particle in an electro-magnetic field}

In the previous section, we analyzed {the situations in which} the equations of motion follow directly from a potential energy due to the velocity-independent nature of the forces involved. We now turn to systems in which the force depends explicitly on the particle's velocity and examine whether the Heisenberg picture reproduces the corresponding classical equations of motion for such systems.

{As an illustrative and physically motivated example, we consider the motion of a charged particle in an external electromagnetic field, where the force depends explicitly on the velocity through its coupling to the magnetic field. The Hamiltonian of the system is given by:
\begin{equation}
\obs H = \frac{1}{2m}\left(\vect{\obs{p}} -
q\,\vect{\obs{A}}\right)^2 -  q\vect{\mathcal{E}}\cdot\vect{\obs{r}}.
\label{eq:ham-mag}
\end{equation}
where $m$ and $q$ denote the mass and charge of the particle, respectively. The magnetic field and the uniform electric field are given by $\vect{\obs{B}}=\vect{\obs{\nabla}}\times\vect{\obs{A}}$ and $\vect{\mathcal{E}}$, with $\vect{\obs{A}}$ denoting the vector potential.} {In general, the magnetic field, the associated vector potential, as well as the electric field, are position-dependent and should therefore be treated as operators. In the present work, however, we assume the electric field to be spatially uniform (position independent), which permits it to be regarded as a classical entity rather than a quantum one.} The velocity operator is defined via the Heisenberg equation of motion:
\begin{equation}
\obs v_i = \frac{d\obs r_i}{dt} = \tfrac{\i}{\hbar} \left[\obs H,\obs r_i\right].
\end{equation}
Using the canonical commutation relations
\begin{equation}
[\obs r_i,\obs p_j]=\i\hbar\,\delta_{ij}, \qquad [\obs r_i,\obs{A_j}] = 0,
\end{equation}
we compute
\begin{align}
[\obs H,\obs r_i]
&= \frac{1}{2m} \left[ (\obs{p}_j-q\obs{A}_j)^2,\obs r_i
\right] 
= -\frac{\i \hbar}{m} (\obs{p}_i-q\obs{A}_i).
\end{align}
Therefore, the velocity operator is
\begin{equation}
\obs v_i = \tfrac{1}{m} \left(\obs{p}_i-q\obs{A}_i\right)
\label{eq:operator-vel}
\end{equation}
It is convenient to define the \emph{kinetic momentum operator}
\begin{equation}
\obs\pi_i \equiv \obs{p}_i-q\obs{A}_i, \qquad \obs v_i = \frac{\obs\pi_i}{m}.
\label{eq:operator-kin-mom}
\end{equation}
The Heisenberg equation for $\obs\pi_i$ reads
\begin{equation}
\frac{d\obs\pi_i}{dt} = \tfrac{\i}{\hbar} \left[\obs H,\obs\pi_i\right].
\end{equation}
Since
\begin{equation}
\obs H= \sum_{i=1}^3\left(\tfrac{1}{2m}\obs\pi_i\obs\pi_i -  q \mathcal{E}_i \obs{r_i}\right),
\end{equation}
we obtain
\begin{equation}
[\obs H,\obs\pi_i] = \tfrac{1}{2m} \left( \obs\pi_j[\obs\pi_j,\obs\pi_i] +[\obs\pi_j,\obs\pi_i]\obs\pi_j \right) - \i \hbar q\mathcal{E}_i .
\end{equation}
Using $[\obs p_i,f(\obs r)] = -\i \hbar\,\partial_i f(\obs r)$,
we compute
\begin{align}
[\obs\pi_i,\obs\pi_j]
&=[\obs{p}_i-q\obs{A}_i,\obs{p}_j-q\obs{A}_j] \nonumber\\
&= \i \hbar q \left(\frac{\partial\obs{A}_j}{\partial\obs{r}_i}-\frac{\partial\obs{A}_i}{\partial\obs{r}_j}\right) 
= \i \hbar q\sum_{k=1}^3\epsilon_{ijk}\obs{B}_k ,
\end{align}
where $\epsilon_{ijk}$ is the Levi-Civita epsilon. Substituting into the Hamiltonian commutator,
\begin{equation}
[\obs H,\obs\pi_i] = - \frac{\i \hbar q}{m} \sum_{j,k}\epsilon_{ijk}\obs\pi_j B_k - \i \hbar q\mathcal{E}_i.
\end{equation}
Therefore,
\begin{align}
m\frac{d\obs v_i}{dt} &= \frac{\i}{\hbar} [\obs H,\obs\pi_i] 
= q\sum_{j,k} \epsilon_{ijk}\obs v_j B_k + q\mathcal{E}_i,
\end{align}
where $\sum_{j,k} \epsilon_{ijk}\obs v_j B_k=(\vect{\obs{v}}\times\vect{\obs{B}})_i$.
In vector form,
\begin{equation}
 m\frac{d\vect{\obs{v}}}{dt} = q\,\vect{\obs{v}}\times \vect{\obs{B}}
+ q\vect{\mathcal{E}}\unity
\label{lorentz}
\end{equation}
Remarkably, the Lorentz force law emerges \emph{exactly} from operator algebra in quantum mechanics.
\section{Rotating frame of reference}

{Let us look for} other situations where velocity dependent forces arise. It is well known in classical mechanics that {in} a rotating frame of reference, Coriolis force and centrifugal force arise as \emph{pseudo forces}. Let us examine this situation in quantum mechanics. Consider a reference frame $\text{S}'$ that rotates with a constant angular velocity $\vect{\omega}$ around the origin of coordinates in the $xy$ plane of frame S. The Hamiltonian as seen by an observer in the rotating frame is given by \cite{pseudoforces}
{\begin{equation}
\obs{H}=\frac{1}{2m}(\vect{\obs{p}}-m\vect{\omega}\times\vect{\obs{r}})^2
- \frac{1}{2}m(\vect{\omega}\times\vect{\obs{r}})^2,
\end{equation}}
The Heisenberg equations for {position components} are given by
\begin{eqnarray}
\frac{d\obs{r}_1}{dt}=\tfrac{1}{m}(\obs{p}_1+m\omega \obs{r}_2),~~~~
\frac{d\obs{r}_2}{dt}=\tfrac{1}{m}(\obs{p}_2-m\omega \obs{r}_1). 
\end{eqnarray}
{The Heisenberg equations for the velocity components, defined as $\obs{v}_i = {d\obs{r}_i}/{dt}$, have the following form}
\begin{eqnarray}
    \frac{d\obs{v}_1}{dt}=2\omega\obs{v}_2+\omega^2\obs{r}_1,~~~~
    \frac{d\obs{v}_2}{dt} =-2\omega\obs{v}_1+\omega^2 \obs{r}_2 .
\end{eqnarray}
Using $\vect{\omega}=\omega\,\hat z$, these equations combine to
\begin{equation}
m\frac{d^2\vect{\obs{r}}}{dt^2}
=-2m\,\vect{\omega}\times\vect{\obs{v}}
-m\vect{\omega}\times(\vect{\omega}\times\vect{\obs{r}}).
\label{rotating}
\end{equation}
The equation presented above is fully quantum mechanical, and is derived from the Heisenberg equations of motion. On the right hand side one can identify 
$-m\vect{\omega}\times(\vect{\omega}\times\vect{\obs{r}})$ with the centrifugal force and $-2m\,\vect{\omega}\times\vect{\obs{v}}$ with the Coriolis force. It is identical to Newton's second law of motion {in a rotating frame of reference}.

\section{Generalized treatment}
Now that we are clear that quantum and classical dynamics {are} governed by the same equations, {with the Heisenberg operators substituting the classical observables,} in a wide variety of situations, we wish to demonstrate it in a general scenario. Consider a system described by the Hamiltonian, expressed as a function of the generalized coordinates and generalized momenta, $ \obs H(\obs{q}_i,\obs{p}_i)$. Heisenberg equations for each of the generalized coordinates and momenta are given by
\begin{equation}
\frac{d\obs{q}_j}{dt} = \frac{\i}{\hbar}[\obs H(\obs{q}_i,\obs{p}_i),\obs{q}_j],~~~~
\frac{d\obs{p}_j}{dt} = \frac{\i}{\hbar}[\obs H(\obs{q}_i,\obs{p}_i),\obs{p}_j].
\label{heisenpq}
\end{equation}
Assuming that $\obs{H}$ can be expanded in a Taylor series in each generalized coordinate and momentum, we can write
\begin{eqnarray}
\obs{H} &=& \sum_{n=0}^{\infty} \frac{(\obs{p}_j-a_j\unity)^n}{n!}  \left.\frac{\partial^n \obs{H}}{\partial\obs{p}_j^n}\right|_{\obs{p}_j=a_j\unity} \nonumber\\
\obs{H} &=& \sum_{n=0}^{\infty} \frac{(\obs{q}_j-b_j\unity)^n}{n!} \left.\frac{\partial^n \obs{H}}{\partial\obs{q}_j^n}\right|_{\obs{q}_j=b_j\unity} ,
\label{Hpq}
\end{eqnarray}
An important issue here may be the ordering of noncommuting operators. It crops up when one wants to quantize a classical Hamiltonian. The popular ordering rules include the Weyl, simplest symmetric, and the Born-Jordan ordering \cite{Bagunu2025}. It has been shown that the Hamiltonians obtained from Weyl, simplest
symmetric, and Born-Jordan quantization all satisfy the required algebra of the quantum equations of
motion \cite{Bagunu2025}. Although in writing (\ref{Hpq}) we use only one pair of canonically conjugate operators at a time, there may be some subtlety regarding ordering involved here. One may mention here that canonical quantization, although the most popular one, is not the only quantization method. There are situation where it {fails, mostly in particle and field theories \cite{Haag}}. An interesting approach which solves the problem in such situations is affine quantization \cite{Fantoni2025}.  

It should be kept in mind that $\left.\frac{\partial^n \obs{H}}{\partial\obs{p}_j^n}\right|_{\obs{p}_j=a_j\unity}$ is still a function of all the $\obs{q}_i$s and all the $\obs{p}_i$s except $\obs{p}_j$. Similarly, $\left.\frac{\partial^n \obs{H}}{\partial\obs{q}_j^n}\right|_{\obs{q}_j=b_j\unity}$ is a function of all the $\obs{p}_i$s and all the $\obs{q}_i$s except $\obs{q}_j$. {The commutators of the position and momentum operators with the Hamiltonian are then given by}
\begin{eqnarray}
\big[\obs H(\obs{q}_i,\obs{p}_i),\obs{q}_j\big]
&=& -\, \i\hbar \sum_{n=0}^{\infty}
\frac{n(\obs{p}_j-a_j\unity)^{n-1}}{n!}\left.
\frac{\partial^n \obs H}{\partial \obs{p}_j^n} \right|_{\obs{p}_j=a_j\unity} ,
\nonumber \\[6pt]
\big[\obs H(\obs{q}_i,\obs{p}_i),\obs{p}_j\big] &=& \, \i\hbar \sum_{n=0}^{\infty} \frac{n(\obs{q}_j-b_j\unity)^{n-1}}{n!} \left. \frac{\partial^n \obs H}{\partial \obs{q}_j^n} \right|_{\obs{q}_j=b_j\unity}.
\label{comutator}
\end{eqnarray}
Here, we use the identities
\begin{eqnarray}
\big[(\obs p_i - a_i\unity)^n,\obs q_j\big]
&=& -\, \i \hbar\, n\,(\obs p_i - a_i\unity)^{n-1}\,\delta_{ij}, \nonumber \\[6pt]
\big[(\obs q_i - b_i\unity)^n,\obs p_j\big]
&=& \;\, \i \hbar\, n\,(\obs q_i - b_i\unity)^{n-1}\,\delta_{ij}.
\end{eqnarray}
Substituting (\ref{comutator}) in (\ref{heisenpq}), we obtain
\begin{eqnarray}
\frac{d\obs{q}_j}{dt} &=& \frac{\partial}{\partial \obs{p}_j}\sum_{n=0}^{\infty} \frac{(\obs{p}_j-a_j\unity)^{n}}{n!} 
\left.\frac{\partial^n \obs{H}}{\partial \obs{p}_j^n}\right|_{\obs{p_j}=a_j\unity}, \nonumber\\
\frac{d\obs{p}_j}{dt} &=& -\frac{\partial}{\partial \obs{q}_j}\sum_{n=0}^{\infty} \frac{(\obs{q}_j-b_j\unity)^{n}}{n!} 
\left.\frac{\partial^n \obs{H}}{\partial \obs{q}_j^n}\right|_{\obs{q}_j=b_j\unity} .
\end{eqnarray}
The above can be written in a more compact form:
\begin{eqnarray}
\frac{d\obs{q}_j}{dt} &=& \frac{\partial \obs{H}}{\partial \obs{p}_j}, ~~~~~~
\frac{d\obs{p}_j}{dt} = -\frac{\partial \obs{H}}{\partial \obs{q}_j} .
\label{Hamilton}
\end{eqnarray}
Remarkably eqns. (\ref{Hamilton}) are a modified form of the Heisenberg equations, but have the exact structure of the classical Hamilton's equations of motion, if one were to substitute the operators by the corresponding classical variables.

{Some explanation on the significance of this result is in order here. It is well known that there is a close analogy between the quantum Heisenberg equations of motion and the classical equations of motion expressed in terms of the Poisson bracket. Classical equations of motion of position or momentum can be obtained from the Heisenberg equations by \emph{replacing} the commutator bracket divided by $\i\hbar$ by the Poisson bracket, suggesting the correspondence rule $\{p,H\} \longleftrightarrow [\obs{p},\obs{H}]/\i\hbar$ \cite{PeresQM}. The same holds for the position observable. The previous analysis indicates that, regarding position and momentum, this is not simply a correspondence but rather an equality in a specific sense. The right-hand sides of equations (\ref{Hamilton}) represent the Poisson bracket of position with the Hamiltonian, and the Poisson bracket of momentum with the Hamiltonian, respectively. The sole distinction is that the Poisson brackets presented here involve quantum operators.}

{This connection also explains why the equations of motion contain all the quantum features even though they are structurally identical to the classical equations. {It should be noted that such a correspondence is only applicable to observables with classical analogues and does not hold for observables that are intrinsically quantum mechanical, such as spin.} The commutation relation of conjugate observables {(\textit{i.e.} whose commutator yields $\i\hbar$)} is what leads to derivatives appearing in (\ref{Hamilton}).}


\section{Ehrenfest theorem}

Till now we have studied the dynamics of quantum observables at the operator level. The values of observables depend on the quantum state. In an arbitrary quantum state we can always define the expectation values of all observables. If we are interested in looking at the experimentally accessible quantities, the expectation values serve the purpose. One can take the expectation value of both sides of equation (\ref{Nxp}) and get
\begin{equation}
m\frac{d^2}{dt^2}\langle \obs{x}\rangle = \frac{d}{dt}\langle \obs{p}\rangle = -\langle V'({\obs{x}})\rangle ,
\label{ehren}
\end{equation}
where the angular brackets denote the expectation value. This is the familiar equation denoting Ehrenfest theorem \cite{Ehrenfest}. It shows that the expectation values of position and momentum \emph{approximately} follow Newton's equations of motion. It is approximate because
$\langle V'({\obs{x}})\rangle \neq V'( \langle{\obs{x}}\rangle)$. 

For a general potential, if we can expand its derivative in a series, $V'({\obs{x}}) \approx \alpha\obs{x} + \beta\obs{x}^2$, then
\begin{eqnarray}
\langle V'({\obs{x}})\rangle &=& \alpha\langle\obs{x}\rangle + \beta\langle{\obs{x}^2}\rangle 
 = \alpha\langle\obs{x}\rangle + \beta\langle{\obs{x}}\rangle^2 + \beta(\Delta x)^2 \nonumber\\
 &=& V'(\langle {\obs{x}}\rangle) + \beta(\Delta x)^2 ,
 \label{Vpav}
\end{eqnarray}
where $\Delta x$ is the uncertainty in position in the given quantum state.
So, if the state is such that the uncertainty in position is small, then
$\langle V'({\obs{x}})\rangle \approx V'( \langle{\obs{x}}\rangle)$, and one can say that the expectation values follow classical dynamics.
For a harmonic oscillator, since $V'(\obs{x})$ is linear in $\obs{x}$, 
$\langle V'({\obs{x}})\rangle = V'( \langle{\obs{x}}\rangle)$, and the position and momentum expectation values exactly follow classical dynamics, for all quantum states. It is straightforward to see that since eqns. (\ref{lorentz}) and (\ref{rotating}) are linear in the observables, both of them would lead to the expectation values of observables exactly following classical dynamics. {However, any nonlinearity creeping into eqn. (\ref{lorentz}), because of some complicated position dependence of the fields, will cause departure from exact classical dynamics.}
\section{Conclusions}
In this work, we have demonstrated that the dynamical equations governing classical and quantum mechanics possess a deeper and more exact correspondence than is usually emphasized. By focusing on the Heisenberg picture, where observable quantities evolve in time, we showed that the equations of motion for quantum operators can be written in a form identical to Newton’s equations of classical mechanics. Remarkably, this equivalence holds exactly at the operator level and does not require any limiting procedure such as large quantum numbers, small $\hbar$, or semiclassical approximations {for both velocity-independent and velocity-dependent forces}.

For a wide class of systems, ranging from particles moving in arbitrary analytic potentials to charged particles in external electromagnetic fields, the Heisenberg equations reproduce the familiar classical laws of motion, including Newton’s second law and the Lorentz force equation. For a quantum system described in a rotating frame of reference, the Coriolis and centrifugal forces also emerge naturally. In the general treatment, the Heisenberg equations reproduce the Hamilton's equations of motion. The absence of $\hbar$ in these operator equations highlights that the distinction between classical and quantum dynamics does not arise from the form of the equations themselves, but rather from the non-commuting nature of quantum observables and the role of the quantum state. This obviously does not imply that values of quantum observables can be inferred from the classical equations of motion. Values of quantum observables involve difficult issues like quantum measurement, and the uncertainty principle governing the values of observable quantities. This is already indicated from the discussion on the Ehrenfest theorem, and how this operator-level equivalence relates to experimentally accessible quantities. {Eqns. (\ref{ehren}) and (\ref{Vpav}) signal that} even the expectation values of observables, which represent averages over many measurements, do not always obey equations of classical dynamics. Deviations from classical behavior emerge due to quantum uncertainties and higher-order moments. Only under specific conditions, such as narrow wave packets or linear forces, do expectation values exactly follow classical trajectories.

Nevertheless, it is an intriguing observation that \emph{at the operator level} the equations of motion in the quantum and classical formalisms are identical. {This suggests that classical and quantum mechanics are unified in their dynamics as far as the form of the equations is concerned, with the primary distinction lying not in the equations of motion, but rather in the non-commutative nature of observables and the interpretation of physical quantities}. This perspective provides a transparent and exact common ground between the two theories and may offer useful insight into longstanding questions concerning the quantum-classical boundary.

{Of course our results in no way imply that classical mechanics has been obtained by the disappearance of $\hbar$. What they do show is that just to obtain Newton's equations of  motion from quantum mechanics, no small $\hbar$ or large action of the system is needed. Our concept of unification of dynamics refers to the fact that identical equations dictate the time evolution of both classical observables and quantum operators.
We believe this may point to a deeper connection between the two theories.
}
\begin{acknowledgments}
ARS acknowledges financial support from the Ministry of Minority Affairs, India, via Maulana Azad National Fellowship (F. 82-27/2019 (SA-III)). TQ acknowledges useful discussions with Bibhash Paul, T.P. Singh and Pankaj Sharan.
\end{acknowledgments}
\vskip 3mm
\textbf{Author contributions:} TQ formulated the problem. ARS and TQ contributed equally in the calculations and the writing of the manuscript.

\vskip 3mm
\textbf{Author declaration: } The authors have no conflicts to disclose.

\end{document}